\documentclass[twocolumn,floatfix,aps,superscriptaddress,prb]{revtex4}
\usepackage{graphicx}
\usepackage{amsmath}
\usepackage[colorlinks=true, citecolor=blue, urlcolor=blue, linkcolor = blue]{hyperref}
\usepackage{amssymb}
\usepackage{bm}
\usepackage{color}
\usepackage{bookmark}
\usepackage{tabularx}
\usepackage{mathtools}
\usepackage{microtype}
\usepackage{cancel}
\usepackage{relsize}
\usepackage{mathrsfs}
\usepackage{ulem}

\usepackage{xcolor}

\begin{document}

\title{First order dipolar phase transition in the Dicke model with infinitely coordinated frustrating interaction}
\author{S. I. Mukhin}
\affiliation{Instituut-Lorentz, Universiteit Leiden, P.O. Box 9506, 2300 RA Leiden, The Netherlands}
\affiliation{Theoretical Physics and Quantum Technologies Department, NUST ``MISIS", Leninski Avenue 4, 119991 Moscow, Russia}
\author{N. V. Gnezdilov}
\affiliation{Instituut-Lorentz, Universiteit Leiden, P.O. Box 9506, 2300 RA Leiden, The Netherlands}

\date{}

\begin{abstract}
We found analytically a first order quantum phase transition in the Cooper pair box array of $N$ low-capacitance Josephson junctions capacitively coupled to a resonant photon in a microwave cavity. The Hamiltonian of the system maps on the extended Dicke Hamiltonian of  $N$ spins one-half with infinitely coordinated antiferromagnetic (frustrating) interaction. This interaction arises from the gauge-invariant coupling of the Josephson junctions phases to the vector potential of the resonant photon field. In $N \gg 1$ semiclassical limit, we found a critical coupling at which ground state of the system switches to the one with a net collective electric dipole moment of the Cooper pair boxes coupled to superradiant equilibrium photonic condensate. This phase transition changes from the first to second order if the frustrating interaction is switched off. A self-consistently `rotating' Holstein-Primakoff representation for the Cartesian components of the total superspin is proposed, that enables to trace both the first and the second order quantum phase transitions in the extended and standard  Dicke models respectively. 
\end{abstract}

\date{\today}
\maketitle
%\tableofcontents

\section{Introduction}

Realization of the equilibrium photonic condensates is of great interest for  the fundamental study of a new states of light strongly coupled to quantum metamaterials \cite{vonDelft, Mukhin, Fistul, Nakamura, Ciuti, Rabl}. In particular, cavity quantum electrodynamics of superconducting qubits is crucial for the quantum computation perspectives \cite{Wallraff,  Wallraff2, Raimond, DiCarlo}.  In quantum optics, e.g. in the cavity QED described by the famous Dicke model \cite{Dicke}, the `no-go' theorems made those perspectives gloomy \cite{Zakowicz, Birula, Keeling}, and only dynamically driven condensates are considered \cite{Kouwenhoven, Zagoskin, Brandes_noneq, Tsai, Nori, Schon}. Nevertheless, it was found that in the equilibrium circuit QED systems the  `no-go' theorems may not hold \cite{Ciuti, Nakamura}. In particular,  an array of capacitively coupled Cooper pair boxes to a resonant cavity was proven to disobey the `no-go' theorem for an equilibrium superradiant quantum phase transition \cite{Ciuti}. Nevertheless, another complication was found in this case, i.e. it was demonstrated \cite{Rabl, Stroud, Stroud2}, that allowance for the gauge invariance with respect to the electromagnetic vector potential of the photon field causes Hamiltonian of the system to map on the extended Dicke model Hamiltonian of (pseudo)spins one-half, adding to the standard Dicke model a frustrating infinitely coordinated antiferromagnetic interaction between the spins.  Lately, a numerical diagonalization results for small clusters with $N$ spins were reported \cite{Rabl}  to behaved differently depending on the parity of the number of spins $N$.

Motivated by the above history of the extended Dicke model exploration, we present in this paper analytic description of the superradiant equilibrium quantum phase transition in the an array of $N\gg1$ Cooper pair boxes strongly coupled to a resonant cavity. The plan of the present paper is as follows. 

First, we reproduce derivation \cite{Rabl, Stroud} of the extended Dicke Hamiltonian with infinitely coordinated antiferromagnetic (frustrating) term. Next, we confirm the absence of the zero modes in the spectrum of the bosonic excitations, as was found in \cite{Rabl}. Then, we introduce a new representation for the operators of Cartesian components of the total spin (`superspin') of  $N$ spins $1/2$: a self-consistently rotating Holstein-Primakoff (RHP) representation. After that, we demonstrate that RHP method applied to extended Dicke Hamiltonian reveals the first order quantum phase transition, that sets the system into a double degenerate dipolar ordered superradiant state with coherent photonic condensate emerging in the cavity. Besides, in Appendix \ref{app_Dicke} we show that the RHP approach also reproduces the second order quantum phase transition for the Dicke Hamiltonian without frustrating interaction term, found earlier by other method \cite{Brandes, Brandes_entlg}. We discuss a drastic difference between the critical values of the coupling strength $g_c$ in the $N \to \infty$ limit  for the 1st and 2nd order phase transitions. In the Summary we present some evaluations of the parameters of a possible experimental setup for a validation of our theoretical predictions for Cooper pair box arrays in a microwave cavity. 

\section{Dicke Hamiltonian for a Cooper pair boxes array} 
 
In this section we present a derivation of the extended Dicke model Hamiltonian. We consider a single mode electromagnetic resonant cavity of a linear dimension $L$ coupled to the array of $N$ independent dissipationless Josephson junctions. It is assumed that the wavelength $\lambda$ of the cavity's resonant photon is much greater than the inter-junction distance: 
$\lambda \gg L/N$. The vector potential of the electromagnetic field related with the photon is expressed in the second quantized form: 
\begin{align}
\vec{A}=\sqrt{\frac{c^2h}{\omega V}}\left( \hat{a}^\dag + \hat{a}\right) \vec{\epsilon}\,, 
 \label{A}
 \end{align}
where $h$ - is Planck's constant, $\omega$ is bare photon frequency, the photon creation and annihilation bosonic operators are $ \hat{a}^\dag,\;\hat{a}$, $\vec{\epsilon}$ is polarization of the electric field, $c$ is velocity of light, and $V$ is the volume of the cavity.

The Hamiltonian of the Cooper pair box array in the cavity then reads:  
\begin{align}
&{}\hat{H}=\hat{H}_{ph}+\hat{H}_{JJ}\, \label{H}, \\
&{}\hat{H}_{ph}=\frac{1}{2}\left(\hat{p}^2+\omega^2 \hat{q}^2 \right) \, \label{H_photon},\\
&{}\hat{H}_{JJ}= E_C \sum_{i=1}^N \hat{n}^2_i - E_J \sum_{i=1}^N \cos\left(\hat{\phi}_i -\displaystyle{\frac{g}{\hbar}} \hat{q}\right) \, \label{H_JJ},
\end{align}
where the coupling constant is  $g={2 el\sqrt{4\pi}}/\sqrt{V}$, and $l$ is of the order of a penetration depth of an electric field  into the superconducting islands forming the Josephson junction, thus, giving the effective thickness across of it \cite{Stroud2}. For simplicity, we consider all the junctions being identical, with electric field polarization $\vec{\epsilon}$ aligned across a Josephson junction. Here the two mutually commuting sets of the conjugate variables are introduced: $\left[\hat{p},\hat{q}\right]=-\mathrm{i\hbar}$ and $\left[\hat{n}_i,\hat{\phi}_i\right]=- \mathrm{i}$. 
The second quantized (harmonic oscillator) variables of the photonic field  are: 
\begin{align}
\hat{p}=\mathrm{i} \sqrt{\frac{\hbar\omega}{2}}\left(\hat{a}^\dag -\hat{a} \right)\;\;\; \text{and} \;\;\;  \hat{q}= \sqrt{\frac{\hbar}{2\omega}}\left(\hat{a}^\dag + \hat{a} \right)\, \label{pq},
\end{align}
where $\left[\hat{a},\hat{a}^\dag\right]=1$.
An operator $2e\hat{n}_i=2e\left(\hat{n}_i^R-\hat{n}_i^L\right)/2$ stands for half of a charge difference at the $i$-th junction, and equals half the difference of the number of Cooper pairs populating the left and right islands of a Josephson junction accordingly, multiplied by the elementary charge $2e$ of the Cooper pair. The quantum of the charging energy of a single junction is $E_C=\left(2 e\right)^2/2 C$.

Following \cite{Stroud} we make a canonical transformation: 
\begin{align}
\hat{\phi}'_i=\hat{\phi}_i -\displaystyle{\frac{g}{\hbar}} \hat{q}\,  \;\;\; \text{and} \;\;\; \hat{n}'_i=\hat{n}_i \, \label{canonical_transform_JJ}
\end{align}
for the JJ's variables and
\begin{align}
\hat{p}'=\hat{p} + g \sum_{i=1}^N \hat{n}_i\,  \;\;\; \text{and} \;\;\; \hat{q}'=\hat{q} \, \label{canonical_transform_pq}
\end{align}
for photonic variables, 
so that $\left[\hat{p}',\hat{\phi}'_i\right]=0$ and the other commutation relations between all the operators remain intact. The Hamiltonian (\ref{H}) becomes
\begin{align} \nonumber
\hat{H} &{}=  \frac{1}{2}\left(\hat{p}^2+\omega^2 \hat{q}^2 \right) - g \hat{p} \sum_{i=1}^N \hat{n}_i +\frac{g^2}{2}  \left(\sum_{i=1}^N \hat{n}_i\right)^2  \\&{} + \sum_{i=1}^N\left( E_C \, \hat{n}_i^2 - E_J \cos \hat{\phi}_i \right) \, \label{H_transformed},
\end{align}
where the primes in the new variables are omitted for brevity. Thus, the infinitely coordinated interaction term $\propto g^2$ has appeared in (\ref{H_transformed}) after the canonical transformation of the Hamiltonian (\ref{H}).

We restrict ourselves to the Cooper pair box limit \cite{Shnirman}, when the charging energy $E_C$ is large in comparison with Josephson coupling $E_J$ and the eigenstates of the Hamiltonian (\ref{H_JJ}) in the zero order approximation can be chosen as the eigenstates of the charge difference operators $\hat{n}_i$. The lowest bare energy level corresponding to the quantum states $\left| n_i=-\frac{1}{2}\right\rangle$ and $\left| n_i=\frac{1}{2}\right\rangle$ is thus twofold degenerate with respect to the direction of the Cooper pair box dipole moment $\vec{d}_i=2el\hat{n}_i\vec{\epsilon}$ ($l$ - is effective thickness of the $i$-th JJ). This double-degenerate level is separated from the levels with the greater charge differences by the $E_C$ gap. The Josephson tunnelling term $\sim E_J$ lifts the degeneracy and opens a gap between the energy levels of the two states that differ by the wave-function parity $\pm 1$ with respect to inversion of the Cooper pair box dipole's direction. Thus, formed two-level system is naturally described by the Pauli matrices $\hat{\sigma}_i^\alpha$. On the subset of these lowest energy states the initial Hamiltonian (\ref{H_JJ}) of the Cooper pair box array of $N$ Josephson junctions is represented by a Hamiltonian of interacting $N$ spins half: 
\begin{align} \nonumber
\hat{H}_{JJ}&{}=\sum_{i=1}^N \left(E_C \, \hat{n}_i^2 - E_J \cos \hat{\phi}_i\right)\approx\\ \nonumber &{}= \frac{E_C}{4} \sum_{i=1}^N\left(\left|{\frac{1}{2}}\right\rangle \left\langle{\frac{1}{2}}\right| + \left|-{\frac{1}{2}}\right\rangle \left\langle -{\frac{1}{2}}\right|\right)_i  \\ \nonumber &{}- \frac{E_J}{2} \sum_{i=1}^N\left( \left|{\frac{1}{2}}\right\rangle \left\langle -{\frac{1}{2}}\right| +\left|-{\frac{1}{2}}\right\rangle \left\langle{\frac{1}{2}}\right| \right)_i = \\ &{}=\frac{N E_C}{4} \, \hat{1}- \frac{E_J}{2}\sum_{i=1}^N \hat{\sigma}^x_i \, \label{H_JJ_Cooper pair box}.
\end{align}
Here charge and phase difference operators $\hat{n}_i$ and $\cos \hat{\phi}_i$ are projected on $\hat{s}_i^z$ and $\hat{s}_i^x$ correspondingly, where $\hat{s}_i^\alpha = \frac{1}{2} \hat{\sigma}_i^\alpha$ are spin-1/2 operators expressed via the Pauli matrices. As a result, initial Hamiltonian (\ref{H_transformed}) reduces to the following spin-boson Hamiltonian, {\it{modulo}} energy shift ${N E_C}/{4}$ :
\begin{align} \nonumber
\hat{H} &{}=  \frac{1}{2}\left(\hat{p}^2+\omega^2 \hat{q}^2 \right) - {g} \hat{p} \sum_{i=1}^N \hat{s}^z_i  \\&{}- E_J\sum_{i=1}^N \hat{s}^x_i +\frac{{g}^2}{2} \left( \sum_{i=1}^N \hat{s}^z_i\right)^2   \label{H_Cooper pair box}.
\end{align}
It is important to clarify here the meaning of the spin-boson interaction term in (\ref{H_Cooper pair box}), that had emerged when canonical transformation (\ref{canonical_transform_JJ}, \ref{canonical_transform_pq}) of the initial gauge-invariant Hamiltonian (\ref{H}) was performed :
\begin{align} 
-{g} \hat{p} \, \hat{s}_i^z &{}= - \mathrm{i}\sqrt{\frac{\hbar\omega}{2 V}} \left(\hat{a}^\dag-\hat{a}\right)  \; 2el\sqrt{4\pi}\hat{s}_i^z  =  - \hat{\vec{\mathscr{E}}} \, \hat{\vec{d}}_i \, \label{dipole_energy} 
\end{align}
and represents the energy of the dipole in the electric field.
The electric field operator in (\ref{dipole_energy}) is given by
\begin{align}
\hat{\vec{\mathscr{E}}}= \mathrm{i} \sqrt{\frac{h\omega}{V}} \left(\hat{a}^\dag-\hat{a}\right) \vec{\epsilon}\, \label{E}
\end{align}
and the dipole moment of the single junction is 
\begin{align}
\hat{\vec{d}}_i =  2 e \hat{s}_i^z l \vec{\epsilon}\, \label{dipole_i}.
\end{align} 
The total dipole moment is then 
\begin{align}
\hat{\vec{d}}=\sum_{i=1}^N \hat{\vec{d}}_i = 2 e \, l\vec{\epsilon}\sum_{i=1}^N \hat{s}_i^z   \label{dipole_total}.
\end{align}

For convenience of the further calculations we perform a unitary transformation $U^\dag \hat{H} U$, where
$ U=\frac{1}{\sqrt{2}}\begin{pmatrix}
1 & \mathrm{i} \\ 1 & -\mathrm{i}
\end{pmatrix}$ that interchanges operators of the Cartesian components of spin half: $\hat{s}_i^z \to -\hat{s}_i^y$, $\hat{s}_i^y \to -\hat{s}_i^x$ and $\hat{s}_i^x \to \hat{s}_i^z$. 

Hence, the final Cooper pair box Hamiltonian, that we are going to explore, becomes:  
\begin{align}
\hat{H} =  \frac{1}{2}\left(\hat{p}^2+\omega^2 \hat{q}^2 \right) + g \hat{p} \, \hat{S}^y  - E_J\, \hat{S}^z +\frac{g^2}{2} \left(\hat{S}^y\right)^2  \, \label{UHU_Cooper pair box},
\end{align}
where we have introduced operators $\hat{S}^\alpha=\sum_i \hat{s}_i^\alpha$ of  the total spin components. The total spin $\hat{S}^2$ is conserved, because it commutes with (\ref{UHU_Cooper pair box}): $\left[\hat{S}^2, H\right]=0$. Cooper pairs tunnelling is represented by 
$-E_J \hat{S}^z$ term, $g\hat{p}\,\hat{S^y}$ is a dipole coupling strength between Cooper pair box and photonic field, and $({g^2}/{2}) \left( \hat{S}^y\right)^2$ stands for the infinitely coordinated `antiferromagnetic' frustrating term.

\section{Diagonalization of the frustrated Dicke model}\label{Dicke_Cooper pair box_model}
\subsection{Tunnelling regime}\label{Tunnelling_regime}
\begin{figure}[h!!]
\centerline{\includegraphics[width=1.\linewidth]{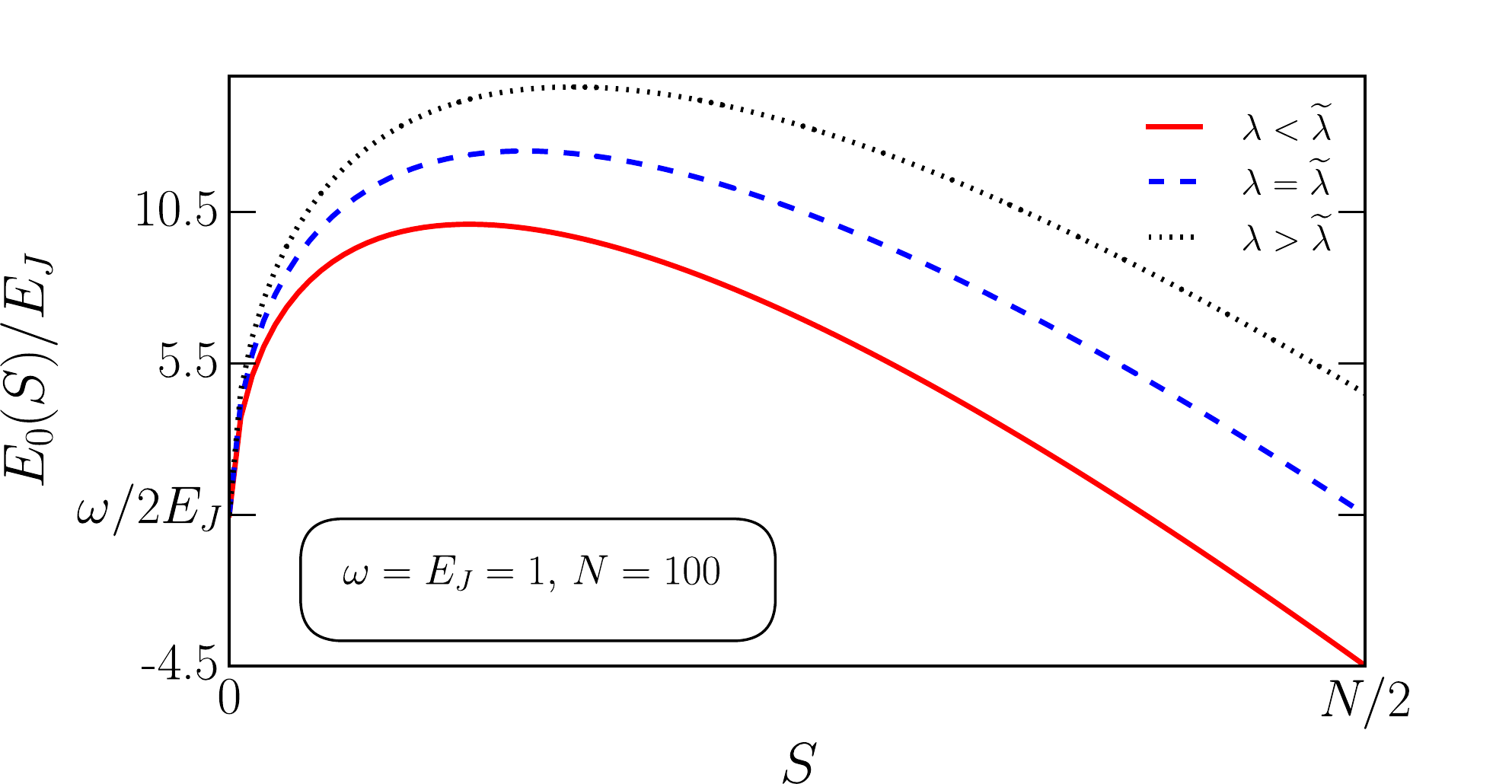}}
\caption{Ground state energy as a function of the superspin $S$ at fixed dimensionless coupling constant $\lambda=g\sqrt{N/2E_J}$. The blue dashed line shows the double degenerate minimum of the ground state at the coupling strength $\lambda=\tilde{\lambda}$  (\ref{g_mf}).}
\label{MF_tunnelling}
\end{figure}

In this section, we consider the frustrated Dicke Hamiltonian (\ref{UHU_Cooper pair box})
and first assume that at small coupling strength $g$ the Josepson tunneling term $-E_J \hat{S}_z$ dominates at zero temperature. Then the  superspin  is in the large $S$ sector and hence, one is allowed to use the Holstein-Primakoff transformation \cite{HP} (HP) in the form:
%\begin{align}
%\begin{cases}\hat{S}^z=S-\hat{b}^\dag \hat{b} \\  
%\hat{S}^y=\mathrm{i}\displaystyle\sqrt{\frac{S}{2}} \left( \hat{b}^\dag \sqrt{1-\frac{\hat{b}^\dag \hat{b}}{2S}} - \sqrt{1-\frac{\hat{b}^\dag \hat{b}}{2S}} \, \hat{b} \right) \simeq \mathrm{i}\displaystyle\sqrt{\frac{S}{2}} \left( \hat{b}^\dag  - \hat{b} \right) \\ 
%\hat{S}^x=\displaystyle\sqrt{\frac{S}{2}} \left( \hat{b}^\dag \sqrt{1-\frac{\hat{b}^\dag \hat{b}}{2S}} + \sqrt{1-\frac{\hat{b}^\dag \hat{b}}{2S}} \, \hat{b} \right) \simeq \sqrt{\frac{S}{2}} \left( \hat{b}^\dag  + \hat{b} \right)\end{cases}  \label{HP},
%\end{align}
\begin{align}
&{}\hat{S}^z=S-\hat{b}^\dag \hat{b}, \label{S_z_HP} \\  
&{}\hat{S}^y=\mathrm{i}\displaystyle\sqrt{\frac{S}{2}} \left( \hat{b}^\dag \sqrt{1-\frac{\hat{b}^\dag \hat{b}}{2S}} - \sqrt{1-\frac{\hat{b}^\dag \hat{b}}{2S}} \, \hat{b} \right) \simeq \mathrm{i}\displaystyle\sqrt{\frac{S}{2}} \left( \hat{b}^\dag  - \hat{b} \right), \label{S_y_HP} \\ 
&{}\hat{S}^x=\displaystyle\sqrt{\frac{S}{2}} \left( \hat{b}^\dag \sqrt{1-\frac{\hat{b}^\dag \hat{b}}{2S}} + \sqrt{1-\frac{\hat{b}^\dag \hat{b}}{2S}} \, \hat{b} \right) \simeq \sqrt{\frac{S}{2}} \left( \hat{b}^\dag  + \hat{b} \right)  \label{S_x_HP},
\end{align}
where $\left[\hat{b},\hat{b}^\dag\right]=1$.
The substitution of (\ref{S_z_HP}, \ref{S_y_HP})  into (\ref{UHU_Cooper pair box})  gives Hamiltonian of the two linearly coupled harmonic oscillators:
\begin{align} \nonumber
\hat{H} &{}=\omega\left( \hat{a}^\dag \hat{a}+\frac{1}{2}\right) -E_J \left(S-\hat{b}^\dag \hat{b}\right) \\{}&- \frac{g\sqrt{S \omega}}{2}  \left(\hat{a}^\dag-\hat{a}\right) \left(\hat{b}^\dag  - \hat{b} \right) - \frac{g^2 S}{4}\left(\hat{b}^\dag  - \hat{b} \right)^2 \,\label{H_ab}. 
\end{align}
This model also arises in the case of ultrastrong light-matter coupling regime with polariton dots \cite{Todorov}.
Here and in what follows we put $\hbar=1$. With the help of the usual linear Bogoliubov's transformation of the creation/annihilation operators (see appendix \ref{app_Cooper pair box}) we obtain diagonalized Hamiltonian:
\begin{align} \nonumber
\hat{H}&{}=-E_J\left(S+\frac{1}{2}\right) + \frac{1}{2}\left(\varepsilon_1+\varepsilon_2\right) \\&{}+ \varepsilon_1\, \hat{c}^\dag_1 \hat{c}_1 + \varepsilon_2 \, \hat{c}^\dag_2 \hat{c}_2 \, \label{H_diag}
\end{align}
with the excitations spectrum described by the new oscillator frequencies:
\begin{align} \nonumber
2 \varepsilon_{1,2}^2 &{}=  E_J\left(E_J+g^2 S\right) +\omega^2   \\ &{} \pm \sqrt{\left(E_J\left(E_J+g^2 S\right)-\omega^2\right)^2 + 4 \omega^2 g^2 S E_J} \, \label{eps_normal},
\end{align}
where the frequencies $\varepsilon_{1,2}$ have to be chosen positive to keep the hermiticity of the initial operators $\hat{p}$, $\hat{q}$, $\hat{S}_y$. 
In contrast with the  Dicke model without frustration \cite{Brandes} both energy branches are real in the whole range of the coupling constants $g$, but with a caveat. Namely, the ground state energy equals:
\begin{align}
E_0(S)=-E_J\left(S+\frac{1}{2}\right) + \frac{1}{2}\left(\varepsilon_1+\varepsilon_2\right) \, \label{E_0}.
\end{align}
This ground state is stable as long as the ground state energy $E_0(S)$ (\ref{E_0}) has global minimum as a function of the superspin $S$ at the end of the interval $\left[0,\,N/2\right]$. One can find value of the coupling strength $g=\tilde{g}$, at which the minimum becomes double degenerate via solving equation $E_0\left(S=N/2, g=\tilde{g}\right)=E_0\left(S=0, g=\tilde{g}\right)=\omega/2$: 
\begin{align}
\tilde{g}\simeq\sqrt{2 E_J N} + \left(E_J+\omega\right) \sqrt{\frac{2}{N E_J}} \, \label{g_mf},
\end{align}
which can be easily derived from large $g$ asymptotic expression of $E_0(S)$:
\begin{align}
E_0 \simeq -E_J\left(S+\frac{1}{2}\right) + \frac{g}{2} \sqrt{E_J S} \label{E0_g}.
\end{align}  
\noindent  For $g > \tilde{g}$ the minimum of $E_0(S)$ migrates from $S=N/2$ to $S=0$, see Figure \ref{MF_tunnelling}. This `jump' of the minimum obviously makes ground state $S=N/2$ unstable and leads to an inapplicability of the quasi-classical HP approximation. Thus, our large $S$ ground state description (\ref{E_0}) is justified for $g<\tilde{g}$. 

\subsection{Rotating Holstein-Primakoff representation} \label{RHP}
In order to make continuation of the theory into the stronger coupling strength region outside the interval $g<\tilde{g}$ , we substitute in the Hamiltonian (\ref{UHU_Cooper pair box}) the $y,z$ components of the total spin operators with a generalised expression of the Holstein-Primakoff representation in a coordinate frame rotated by an angle $\theta$ in the $z$-$y$ plane : 
\begin{align}
\begin{cases} \hat{S}^z=\hat{J}^z \cos \theta - \hat{J}^y \sin \theta \\ \hat{S}^y=\hat{J}^z \sin \theta + \hat{J}^y \cos \theta  \end{cases}\, \label{J}
\end{align}
Here the set of operators of the Cartesian projections of the total spin $\hat{J}^{x,y,z}$ are 
\begin{align}
\begin{cases}\hat{J}^z=S-\hat{b}^\dag \hat{b} \\  
\hat{J}^y \simeq \mathrm{i}\displaystyle\sqrt{\frac{S}{2}} \left( \hat{b}^\dag  - \hat{b} \right) \\ 
\hat{J}^x\simeq \sqrt{\frac{S}{2}} \left( \hat{b}^\dag  + \hat{b} \right)\end{cases} \, \label{HP_J}.
\end{align}
To find $\theta\neq 0$ solution that diagonalises (\ref{UHU_Cooper pair box}) we  introduce a shift, $i\,\sqrt{\alpha}$, of the photon creation/annihilation operators, similar to \cite{Brandes}, in the following way:
\begin{align}
\begin{cases}\hat{a}^\dag = \hat{c}^\dag - \mathrm{i}\sqrt{\alpha} \\ \hat{a} = \hat{c} + \mathrm{i}\sqrt{\alpha} \end{cases}\, \label{a_shift},
\end{align}
\noindent thus, envisaging formation of a superradiant state.
One substitutes (\ref{J} - \ref{a_shift}) into (\ref{UHU_Cooper pair box}) and the Hamiltonian, quadratic in operators $c,c^\dag\,b,\,b^\dag$,  becomes:
\begin{align}\nonumber
\hat{H}&{}=\omega\left(\hat{c}^\dag \hat{c} +\frac{1}{2} \right) - E_J\cos\theta \left(S-\hat{b}^\dag \hat{b}\right) \\&{} - \frac{g \cos \theta \sqrt{S \omega}}{2} \left(\hat{c}^\dag-\hat{c}\right)\left(\hat{b}^\dag - \hat{b}\right)  - \frac{g^2 \cos^2 \theta \, S}{4}\left(\hat{b}^\dag - \hat{b}\right)^2  \label{H1_mu},
\end{align}
\noindent
where an elimination of the linear in $\left(\hat{c}^\dag-\hat{c}\right)$ and $\left(\hat{b}^\dag-\hat{b}\right)$ terms in the Hamiltonian introduces a system of the two equations:
\begin{align}
&\sqrt{2 \omega \alpha}+g \sin \theta\left( S - \left\langle\hat{b}^\dag \hat{b}\right\rangle \right) =0 \,;  \label{c_lin}\\ \nonumber
&E_J \sin \theta + g \cos \theta \sqrt{2 \omega \alpha} \\&{}\;\;\;\;\;\;\;\;\;\;\;\;+ g^2 \cos \theta \sin \theta \left( S - \left\langle\hat{b}^\dag \hat{b}\right\rangle - \frac{1}{2}\right)=0 \, \label{d_lin}.
\end{align}
\noindent We have also made in (\ref{H1_mu}) a mean-field decoupling of the products that are higher than quadratic in $b,\,b^\dag$-operators: $\hat{b}^\dag \hat{b} \, \hat{b}^\dag \hat{b}= 2 \left\langle \hat{b}^\dag \hat{b}\right\rangle  \hat{b}^\dag \hat{b} -\left\langle \hat{b}^\dag \hat{b}\right\rangle^2$ and $\hat{b}^\dag \hat{b} \left( \hat{b}^\dag - \hat{b}\right) + \left( \hat{b}^\dag - \hat{b}\right) \hat{b}^\dag \hat{b} = \left(\hat{b}^\dag - \hat{b}\right)  \left(1+2\left\langle\hat{b}^\dag \hat{b}\right\rangle \right)$.
%\begin{align}
%\hat{b}^\dag \hat{b} \, \hat{b}^\dag \hat{b}= 2 \left\langle \hat{b}^\dag \hat{b}\right\rangle  \hat{b}^\dag \hat{b} -\left\langle \hat{b}^\dag \hat{b}\right\rangle^2 \, \label{d4}
%\end{align}
%and 
%\begin{align}
%\hat{b}^\dag \hat{b} \left( \hat{b}^\dag - \hat{b}\right) + \left( \hat{b}^\dag - \hat{b}\right) \hat{b}^\dag \hat{b} = \left(\hat{b}^\dag - \hat{b}\right)  \left(1+2\left\langle\hat{b}^\dag \hat{b}\right\rangle \right) \, \label{d3}.
%\end{align}

A nontrivial solution $\alpha\neq 0,\,\theta\neq 0$ of  the system of equations (\ref{c_lin}) and (\ref{d_lin}) emerges when $g \ge \sqrt{2 E_J}$ : 
\begin{align}
&{}\cos \theta = \frac{2 E_J}{g^2}\, \label{cos}, \\ \nonumber
&{} \sqrt{\alpha} =  -\frac{g S}{\sqrt{2 \omega}} \left(1-\frac{ \left\langle\hat{b}^\dag \hat{b}\right\rangle}{S} \right) \sin \theta \simeq -\frac{g S}{\sqrt{2 \omega}} \sin \theta =\\&{}\;\;\;\;\;= \,\mp \frac{g S}{\sqrt{2 \omega}} \sqrt{1-\frac{4 E_J^2}{g^4}} \label{alpha05}.
\end{align}

Under the solutions (\ref{cos}), (\ref{alpha05}), the energy of the photonic condensate $=\omega\alpha$ exactly cancels with the sum of the rest of the c-number terms in the Hamiltonian (\ref{H1_mu}): 
\begin{align} \nonumber
&{}\omega \alpha + g S \sin \theta \, \sqrt{2 \omega \alpha} + \frac{g^2 \sin^2\theta}{2}\left(S^2 -\left\langle \hat{b}^\dag \hat{b} \right\rangle^2 \right)= \\ \nonumber &{}=
\frac{g^2 \sin^2 \theta}{2} {\bigg(}S^2 - 2 S \left\langle \hat{b}^\dag \hat{b} \right\rangle+ \left\langle \hat{b}^\dag \hat{b} \right\rangle^2  -2 S^2 \\&{}+ 2 S \left\langle \hat{b}^\dag \hat{b} \right\rangle + S^2 -\left\langle \hat{b}^\dag \hat{b} \right\rangle^2 {\bigg )}=0 \, \label{const1}.
\end{align}
The c-number terms in the first line of (\ref{const1}) have the following meaning: the photonic condensate energy $\sim \omega{\alpha}$,  the (negative) contribution of the dipole-photon coupling energy $\sim g\langle\hat{p}\rangle\langle\hat{S}^y\rangle$, and the zero-point oscillations energy of the frustration term $\sim g^2\langle(\hat{S}^y)^2\rangle/2$. The total of these three terms proves to be zero.  This $\alpha$-independent cancellation, actually, stems from the degeneracy of the energy minima of the diagonal in spin operators part of the extended Dicke Hamiltonian (\ref{UHU_Cooper pair box}) with respect to $2S+1$ different $\hat{S}^y$ projections and classical part $\sim\sqrt{\alpha}$ of the photonic operator $\hat{p}$. 
 
\subsection{Superradiant dipolar regime}\label{Dipolar_regime} 
\begin{figure}[h!!]
\centerline{\includegraphics[width=1.\linewidth]{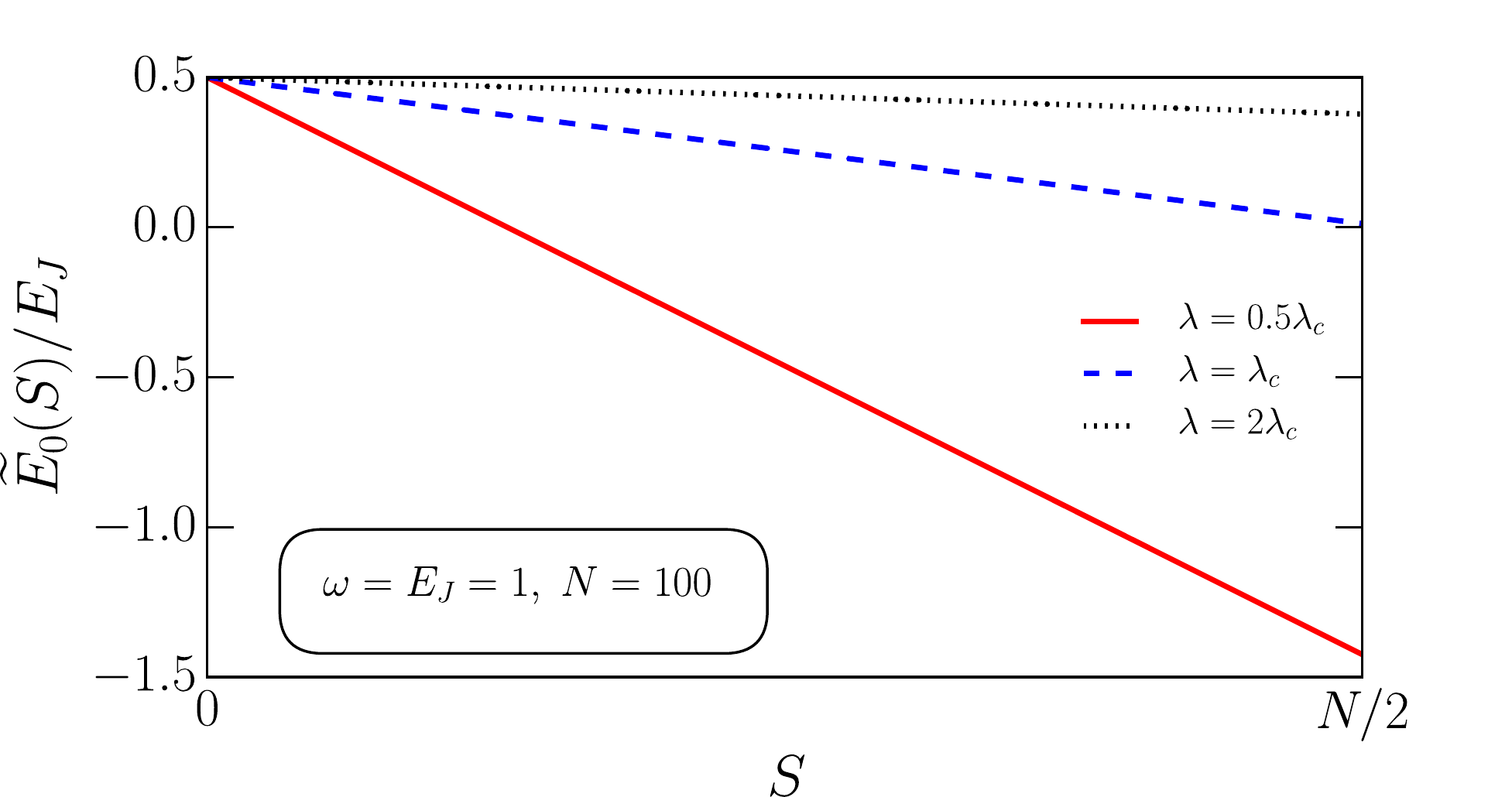}}
\caption{Ground state energy $\widetilde{E}_0$  as a function of the superspin $S$ at fixed dimensionless coupling constant $\lambda=g\sqrt{N/2E_J}$. }
\label{MF_dipole}
\end{figure}

The structure of (\ref{H1_mu}) is the same as (\ref{H_ab}), though with coefficients renormalised with prefactor $\cos\theta$ due to RHP rotation by an angle $\theta$ . Hence, after the Bogoliubov's transformation similar to the one already described in the Appendix \ref{app_Cooper pair box}, the diagonalized Hamiltonian expressed via new second quantized operators $\hat{e}^\dag_{1,2}$, $\hat{e}_{1,2}$ acquires the form:
\begin{align} \nonumber
\hat{H} &{}= -\frac{g^2 \cos^2 \theta}{2} \left(S+\frac{1}{2}\right) + \frac{1}{2} \left(\tilde{\varepsilon}_1 + \tilde{\varepsilon}_2 \right)  \\&{}+\tilde{\varepsilon}_1 \, \hat{e}^\dag_1 \hat{e}_1 +\tilde{\varepsilon}_2 \, \hat{e}^\dag_2 \hat{e}_2\, \label{H1_diag}
\end{align}
with the positive eigenvalues $\tilde{\varepsilon}_{1,2}$
\begin{align} \nonumber
2 \tilde{\varepsilon}^{\,2}_{1,2}&{}= \frac{g^4 \cos^4 \theta}{2}\left(S+\frac{1}{2}\right) +\omega^2 \\&{}\pm \sqrt{\left(\frac{g^4 \cos^4 \theta}{2}\left(S+\frac{1}{2}\right)-\omega^2\right)^2 + 2 S \,\omega^2 g^4 \cos^4 \theta } \, \label{eps_sr}
\end{align}
and the ground state energy 
\begin{align}
\widetilde{E}_0(S) = -\frac{g^2 \cos^2 \theta}{2} \left(S+\frac{1}{2}\right) + \frac{1}{2} \left(\tilde{\varepsilon}_1 + \tilde{\varepsilon}_2 \right) \, \label{E0_tilde}.
\end{align}
The stability of the large $S$ state in this regime is provided by the negative slope of $\widetilde{E}_0(S)$ as a function of $S$ (see Figure \ref{MF_dipole}) in the strong coupling limit:
\begin{align}
\left.\widetilde{E}_0(S)\right|_{g\to+\infty} \simeq \frac{\omega}{2} - \frac{2 E^2_J}{g^2}\left(S+\frac{1}{2}\right) \, \label{E0_tilde_largeg}.
\end{align} 

\noindent
The interval $g \ge \sqrt{2 E_J}$ is characterized with an emergent dipole moment in the Cooper pair box array and the superradint photoinic condensate either as a metastable state for $ \sqrt{2 E_J}\leq g<g_c$, or as the ground state for $g\geq g_c$ (the critical strength $g_c$ is found below).  
To see this explicitly, we express electromagnetic field operators via the new set of Bose-operators found after the Bogoliubov's transformation:
\begin{align} 
&{}\hat{p} = \sqrt{2 \omega \alpha} + \mathrm{i}  \frac{\omega \cos\delta}{\sqrt{2 \tilde{\varepsilon}_1}}    \left(\hat{e}^\dag_1-\hat{e}_1\right) +  \mathrm{i} \frac{\omega \sin\delta}{\sqrt{2 \tilde{\varepsilon}_2}}   \left(\hat{e}^\dag_2-\hat{e}_2\right) \, \label{p_e1e2}, \\ 
&{} \hat{q} =  \frac{\cos\delta}{\sqrt{2 \tilde{\varepsilon}}_1}    \left(\hat{e}^\dag_1+\hat{e}_1\right) + \frac{\sin\delta}{\sqrt{2 \tilde{\varepsilon}}_2}  \left(\hat{e}^\dag_2+\hat{e}_2\right) \, \label{q_e1e2}.
\end{align}
In turn, the spin operators are expressed via $\hat{e}_i,\,{\hat{e}^\dag}_i,\,i=1,2$ as well:
\begin{align}
&{}\hat{J}_z = S \left(1 - \frac{\left\langle \hat{b}^\dag \hat{b}\right\rangle}{S}\right) \simeq S \, \label{JZ}, \\
&{}\hat{J}_y =  - \mathrm{i} \frac{E_J\sqrt{S} \sin\delta}{g\sqrt{\tilde{\varepsilon}_1}} \left(\hat{e}^\dag_1-\hat{e}_1\right) + \mathrm{i}\frac{E_J \sqrt{S} \cos \delta}{g \sqrt{\tilde{\varepsilon}_2}} \left(\hat{e}^\dag_2-\hat{e}_2\right) \label{Jy}, \\ 
&{}\hat{J}_x =  - \frac{E_J\sqrt{S} \sin\delta}{g\sqrt{\tilde{\varepsilon}_1}} \left(\hat{e}^\dag_1+\hat{e}_1\right) +\frac{E_J \sqrt{S} \cos \delta}{g \sqrt{\tilde{\varepsilon}_2}} \left(\hat{e}^\dag_2+\hat{e}_2\right)  \label{Jx}. 
\end{align}
The Bogoliubov's `angle' $\delta$ can be found from the consistency relation:
\begin{align}
\tan 2\delta = \frac{2\sqrt{2 S} \, \omega \, g^2 \cos^2 \theta}{g^4 \cos^4 \theta\left(S+\frac{1}{2}\right)-2\omega^2} \, \label{tan_shift}.
\end{align}

We find for $g\ge \sqrt{2 E_J}$  the following non-zero expectation values in the ground state of Hamiltonian (\ref{H1_diag}). For the electric field  $\hat{\vec{\mathscr{E}}}$:
\begin{align}
&\left\langle\hat{\vec{\mathscr{E}}}\cdot\vec{\epsilon}\right\rangle\sqrt{\dfrac{V}{4\pi}}= \left\langle\hat{p}\right\rangle = \sqrt{2 \omega \alpha} \simeq - g S \sin \theta = \nonumber\\
&=\mp g S \sqrt{1-\frac{4 E_J^2}{g^4}} \,;\label{p_av}
\end{align}
for the modulus of the Josephson tunnelling energy of the Cooper pairs (it decreases): 
\begin{align}
- E_J \langle \hat{S}^z \rangle = -E_J\langle \hat{J}^z \rangle \cos \theta \simeq \, -S \, \frac{2 E_J^2}{g^2} \,; \label{EJ_av}
\end{align}
and a for the emergent finite mean value of the dipole moment:
\begin{align}
\langle \hat{d}\rangle = 2 e l \langle \hat{S}^y \rangle = 2 e l \langle \hat{J}^z \rangle \sin \theta \simeq \pm 2 e l S \sqrt{1-\frac{4 E_J^2}{g^4}}\,. \label{d_av} 
\end{align}  

\begin{figure}[h!!]
\centerline{\includegraphics[width=1.\linewidth]{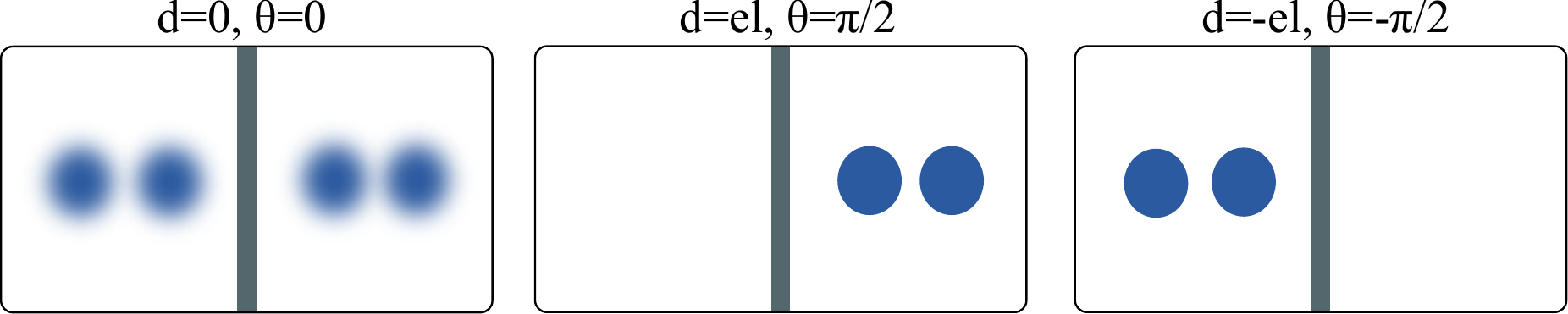}}
\caption{Schematic layout of the amplitude distributions of the Cooper pair's wave function in the adjacent islands of a single JJ and corresponding   dipole moment values depending on the rotation angle $\theta$, see text and equations (\ref{J}).}
\label{domino}
\end{figure}

Hence, results (\ref{p_av}) and (\ref{d_av}) indicate that upon an increase of the coupling strength $g> \sqrt{2 E_J}$  there is a state with the energy given in (\ref{E0_tilde}), which is characterized  by an emergent superradiant electromagnetic field $\left\langle \hat{p}\right\rangle \neq 0$ in the cavity  together with a finite dipole moment of the Cooper pair boxes: $\langle \hat{d}\rangle \neq 0$. The latter means that rotation angle $\theta$ introduced in (\ref{J}) regulates an extent of a Cooper pair wave function between the superconducting islands forming each Josephson junction in the Josephson junction array, see Figure \ref{domino}. Namely, when $\theta$ progressively deviates from zero, the Cooper pairs become localized in one of the two superconducting islands constituting a given Josephson junction, and as a result, the latter acquires a dipole moment.

\section{First order dipolar phase transition}

In this section we calculate a critical coupling $g_c$, at which a first order phase transition between the tunnelling and dipolar states described in Sections \ref{Tunnelling_regime} and \ref{Dipolar_regime} takes place.

\begin{figure}[h!!]
\centerline{\includegraphics[width=1.\linewidth]{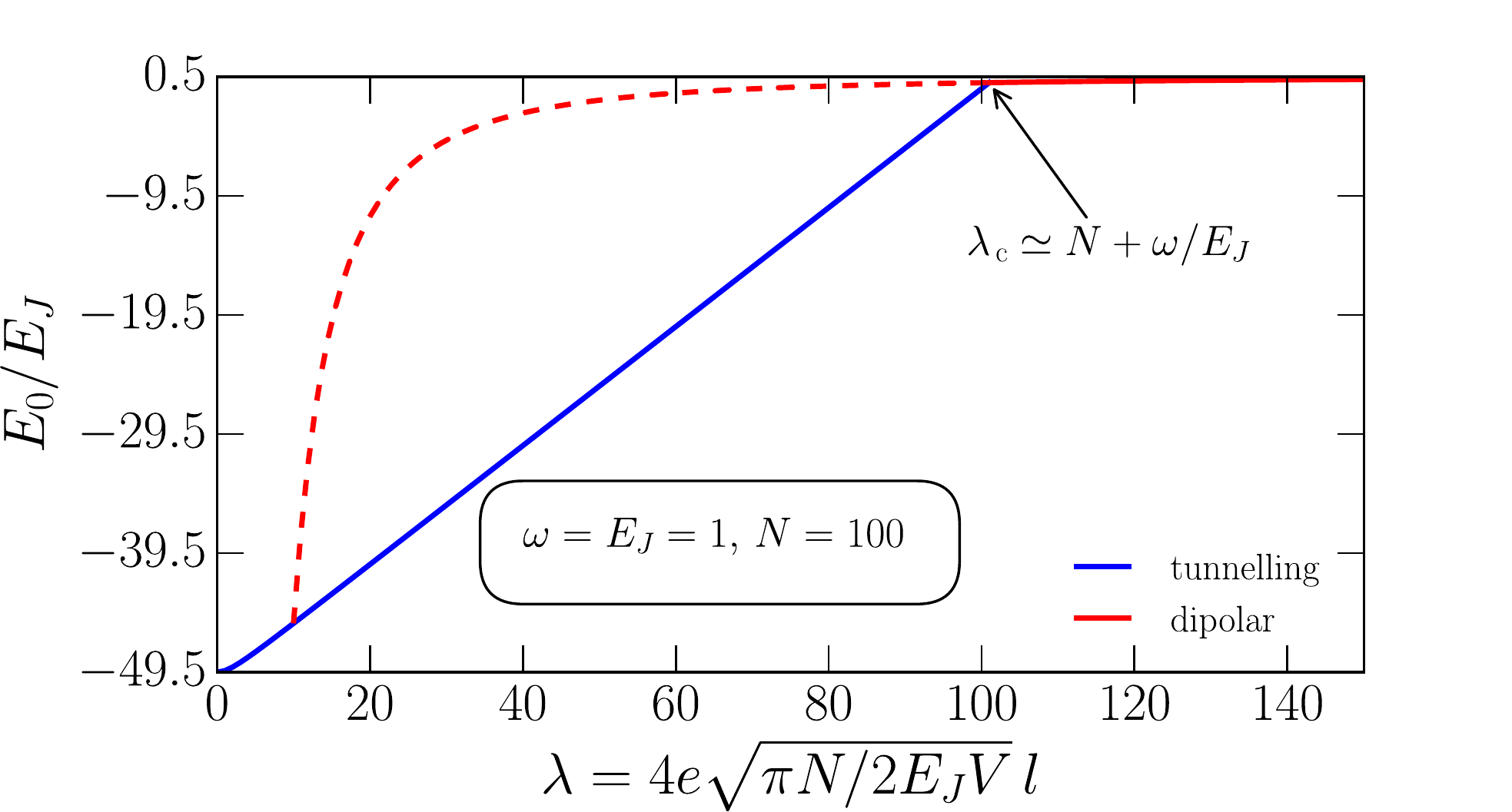}}
\caption{Ground state energy as a function of dimensionless coupling constant $\lambda=g\sqrt{N/2E_J}$. Blue line is for the Josephson tunnelling state in the interval $\lambda< \lambda_\mathrm{c}\approx N$. Red line is for dipolar ordered state. Red dashed line shows the dipolar state in the metastable region preceding the first order phase transition at $\lambda_\mathrm{c}$.}
\label{I_trans}
\end{figure}

In Figure \ref{I_trans}  we plotted ground state energies calculated for tunnelling and dipolar states  as functions of coupling $g$: ${E}_0(S)$ and $\widetilde{E}_0(S)$, see (\ref{E_0}) and (\ref{E0_tilde}) correspondingly. A dimensionless coupling constant $\lambda=g\sqrt{N/2E_J}$ is used. 
In the strong coupling limit, $g \gg \sqrt{2E_J}$, the $g$ dependence of the both branches of energy is very well approximated by (\ref{E0_g}) and (\ref{E0_tilde_largeg}).

Hence, in the thermodynamic limit $N \to \infty$, solution of $E_0=\widetilde{E}_0$ gives the critical value $g_\mathrm{c}$ of the coupling constant: 
\begin{align}g_\mathrm{c} \simeq \sqrt{2 E_J N} + \omega \sqrt{\frac{2}{N E_J}}\, \label{gc}.
\end{align}
\noindent
Here a crucial difference with respect to \cite{Brandes} is that the critical point corresponds to $\lambda_\mathrm{c}\approx N$ and not $1$ as in the standard Dicke model without frustration. Hence, transition now is size-dependent, where the `size' of the system is the total number $N$ of Cooper pair box's inside the microwave cavity.

\begin{figure}[h!!]
\centerline{\includegraphics[width=1. \linewidth]{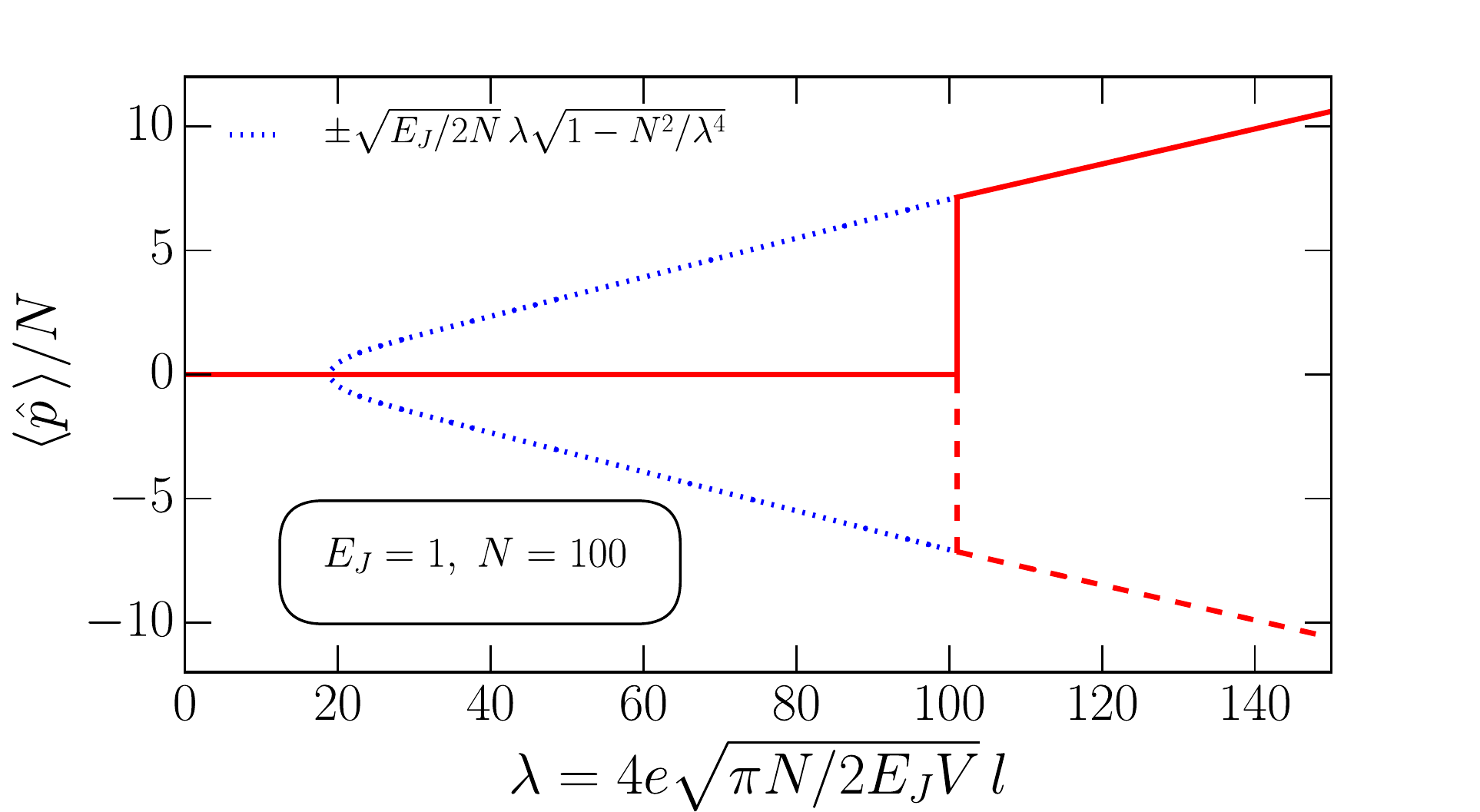}}
\caption{Photon field  $\left\langle \hat{p}\right\rangle$ emerging in the cavity as a function of dimensionless coupling constant $\lambda=g\sqrt{N/2E_J}$. The first order transition to the state with the macroscopic photon occupation number $\left\langle \hat{a}^\dag \hat{a}\right\rangle \neq 0$ occurs at the critical coupling $\lambda_\mathrm{c} \simeq N+\omega/E_J$. The blue dotted line shows the metastable solution for $\left\langle \hat{p}\right\rangle$, that appears at $\lambda=\sqrt{N}$. }
\label{p}
\end{figure}

\begin{figure}[h!!]
\centerline{\includegraphics[width=1. \linewidth]{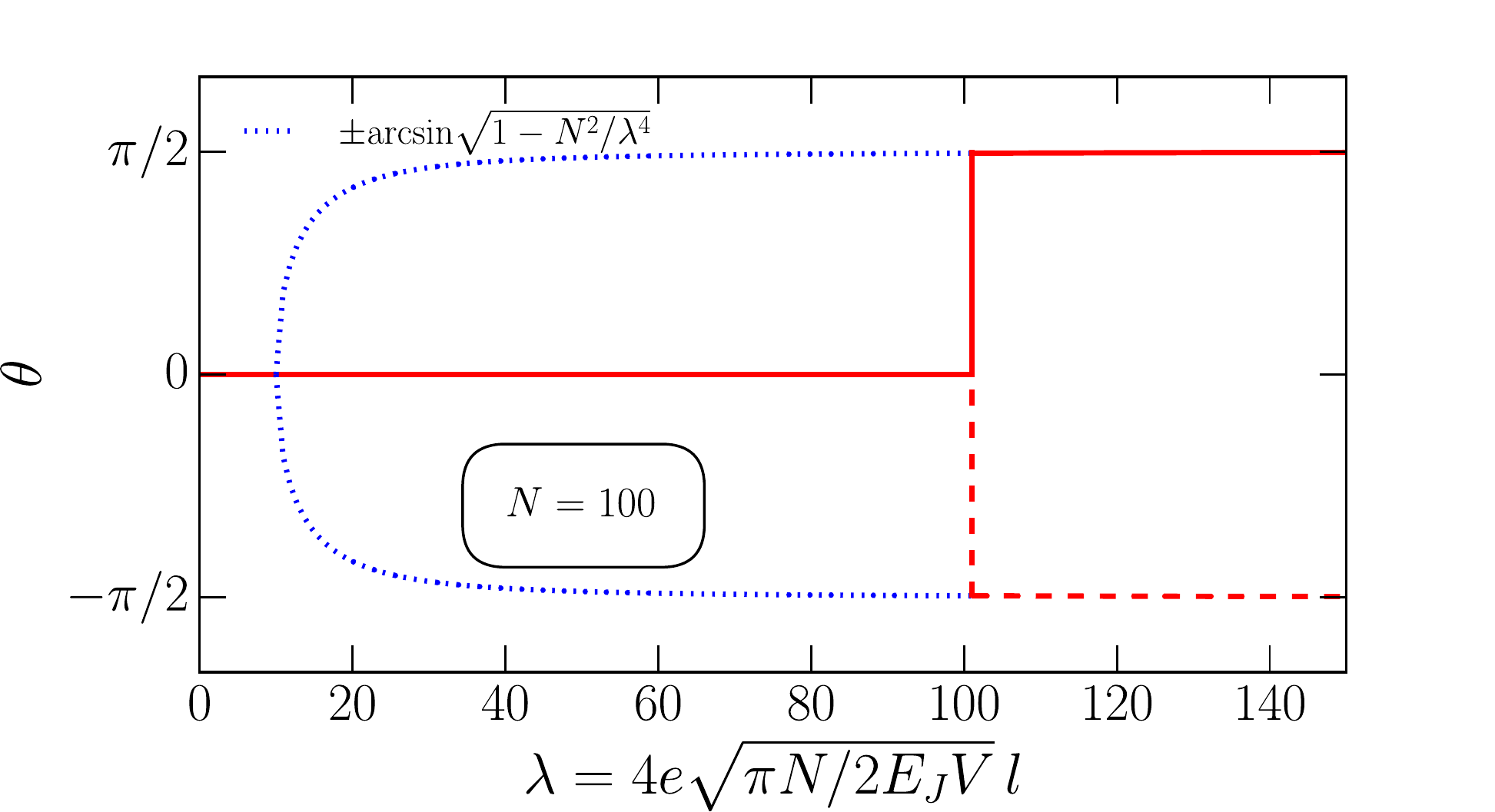}}
\caption{The angle $\theta$, that characterizes rotation of HP, as a function of dimensionless coupling constant $\lambda=g\sqrt{N/2E_J}$. The colour scheme is chosen the same as for the Figure \ref{p}.}
\label{theta}
\end{figure}

\noindent
At $\lambda=\sqrt{N}$, i.e. $g=\sqrt{2E_J}$, the ground state becomes degenerate and a dipolar branch $\widetilde{E}_0(S)$ first appears. For $\sqrt{N}<\lambda<N$, i.e. $\sqrt{2E_J}<g<g_c$, the dipolar state minimal energy $\widetilde{E}_0(S)$ is higher then the tunnelling ground state energy ${E}_0(S)$. Hence, the system remains in the tunnelling state (i.e. dipolar disordered). At $\lambda=\lambda_\mathrm{c}$ the ground state energy ${E}_0(S)$ crosses the dipole state energy branch $\widetilde{E}_0(S)$ for the second time and goes above $\widetilde{E}_0(S)$. 
At the critical coupling $g=g_\mathrm{c}$ (i.e. $\lambda=\lambda_\mathrm{c}$) the first order phase transition from the tunnelling state to dipolar ordered state takes place. It is, indeed, a first order transition, since at $g=g_\mathrm{c}$ dipole moment in the dipolar state is already finite: $\langle \hat{d}\rangle\approx \pm 2 e l S $, see (\ref{d_av}), while in the tunnelling state it equals zero.
Namely, the first order phase transition results in 
\begin{align}
\left\langle \hat{p} \right\rangle = -{Sg}\sin \theta =  \begin{cases} 0\,, \;\;\;  g < g_\mathrm{c} \\ \mp Sg  \sqrt{1-4 E_J^2/g^4}, \;\;\; g \ge g_\mathrm{c} \end{cases} \, \label{p_av_pht}
\end{align}
see Figure \ref{p},
and
\begin{align}
- E_J &{} {\left\langle \hat{S}^z \right\rangle} = -{SE_J} \cos\theta =\begin{cases} -SE_J, \;\;\; g < g_\mathrm{c} \\  -E^2_J2S/g^2, \;\;\; g \ge g_\mathrm{c} \end{cases}  \label{SzS}, \\ 
&{}{\left\langle \hat{d} \right\rangle} =  2e lS\sin \theta = \begin{cases} 0\,, \;\;\; g < g_\mathrm{c} \\ \pm 2e lS \sqrt{1-4 E_J^2/g^4} , \;\;\; g \ge g_\mathrm{c} \end{cases}  \label{SyS}.
\end{align}
The collective dipole moment (\ref{SyS}) is defined by the angle $\theta$, which is shown in the Figure \ref{theta}.

It is important to mention here, that, comparison of (\ref{gc}) with (\ref{g_mf}) gives: $\tilde{g}-g_\mathrm{c}=\sqrt{2 E_J/N} >0$. Hence, we have found the first order phase transition in the region of validity (i.e. $g<\tilde{g}$) of the large superspin limit $S=N/2\gg 1$, that justifies the use of the HP approach. In the limit $g\to +\infty$ the dipolar ordered ground state energy $\widetilde{E}_0$  approaches from below the ground state energy of a free resonant photon, $\omega/2$. Simultaneously, at $g=g_\mathrm{c}$ the ground state energy: $\widetilde{E}_0(S)=0<\omega/2$. Hence, our semiclassical description indicates that after the dipole transition the system gradually approaches decoupled state $\widetilde{E}_0(S)=\omega/2$, but with saturated value of the collective dipole moment $\propto \langle\hat{S}^y\rangle\rightarrow N/2$. It is not possible to decide in the framework of our semiclassical approach whether a crossover  to a state $\langle\hat{S}^y\rangle = 0$ happens in the $g\to +\infty$ limit. The latter state was predicted numerically in finite, even $N$ spin-$1/2$ cluster realization of the extended Dicke model \cite{Rabl}.

\begin{figure}[h!!]
\centerline{\includegraphics[width=1.\linewidth]{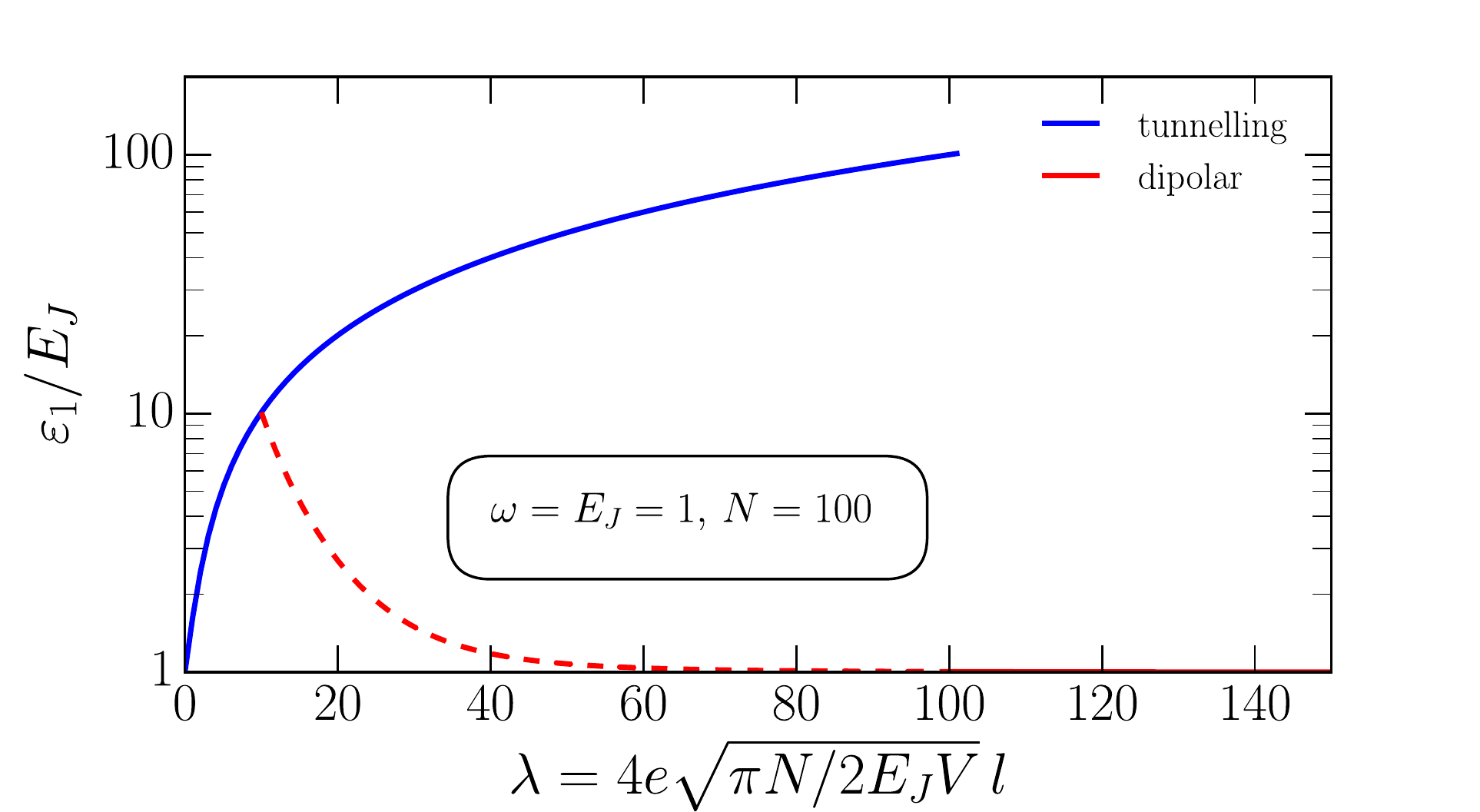}}
\caption{Excitation branches $\varepsilon_1$, $\tilde{\varepsilon}_1$ (\ref{eps_normal}, \ref{eps_sr}) as the functions of dimensionless coupling constant $\lambda=g\sqrt{N/2E_J}$. The vertical axis is shown in the logarithmic scale. At the critical coupling $\lambda_\mathrm{c} \approx N$ the frequency $\varepsilon_1$ jumps to $\tilde{\varepsilon}_1\approx \omega$. The color scheme is the same as in Figure \ref{I_trans}.}
\label{e1}
\end{figure}

\begin{figure}[h!!]
\centerline{\includegraphics[width=1.\linewidth]{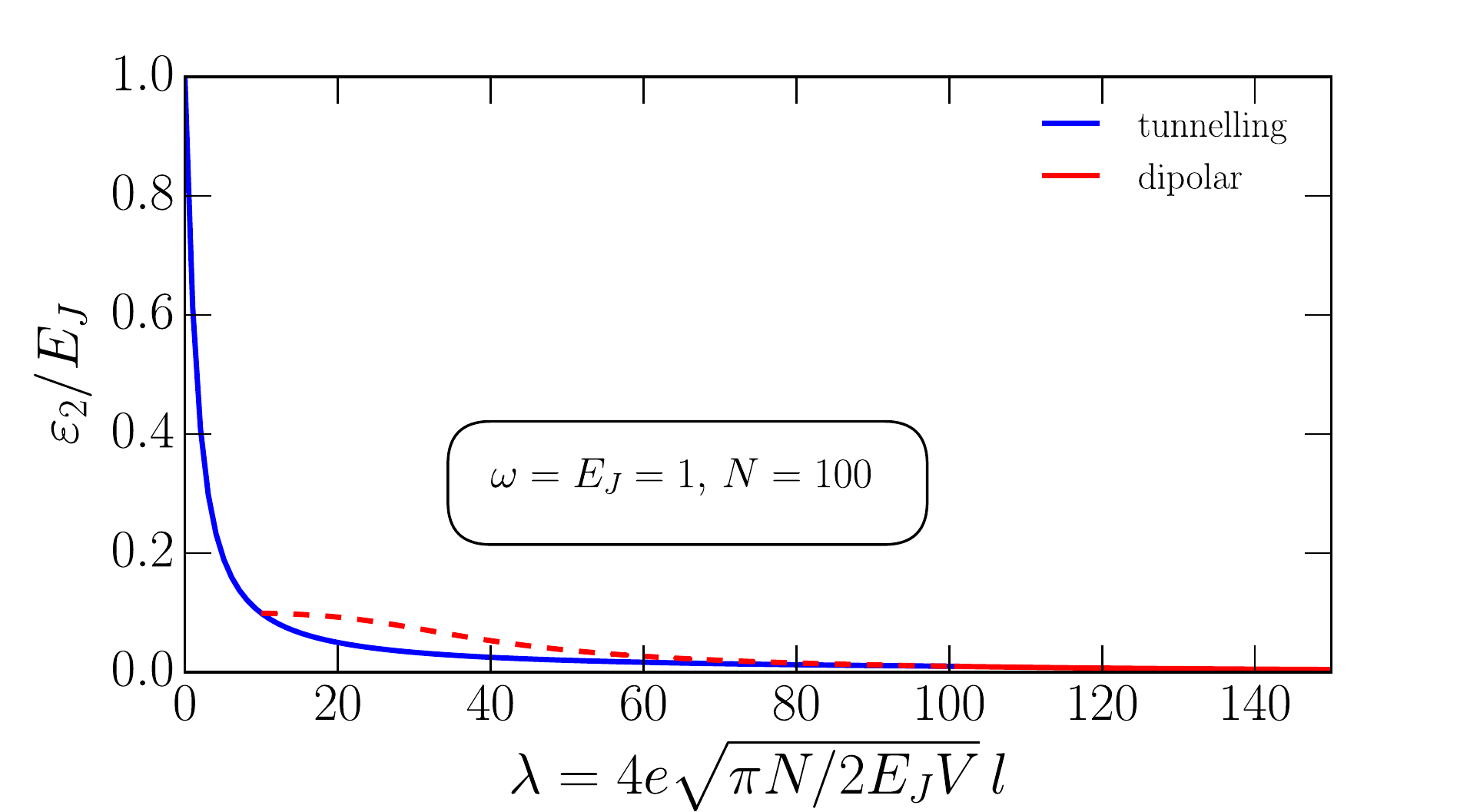}}
\caption{Excitation branches $\varepsilon_2$, $\tilde{\varepsilon}_2$ (\ref{eps_normal}, \ref{eps_sr}) as a function of dimensionless coupling constant $\lambda=g\sqrt{N/2E_J}$. The frequencies $\varepsilon_2$, $\tilde{\varepsilon}_2$ asymptotically approach zero in the strong coupling limit. The color scheme is the same as in Figure \ref{I_trans}.}
\label{e2}
\end{figure}

The excitations branches (\ref{eps_sr}) of the diagonalized Hamiltonian are shown on the Figures \ref{e1}, \ref{e2}. 

\noindent 
The branch $\varepsilon_1$, that grows with the increase of the coupling, goes to the initial photon's frequency $\omega$ after the 
first order transition. $\varepsilon_2$ approaches zero in the strong coupling limit.

\begin{figure}[h!!]
\centerline{\includegraphics[width=1.\linewidth]{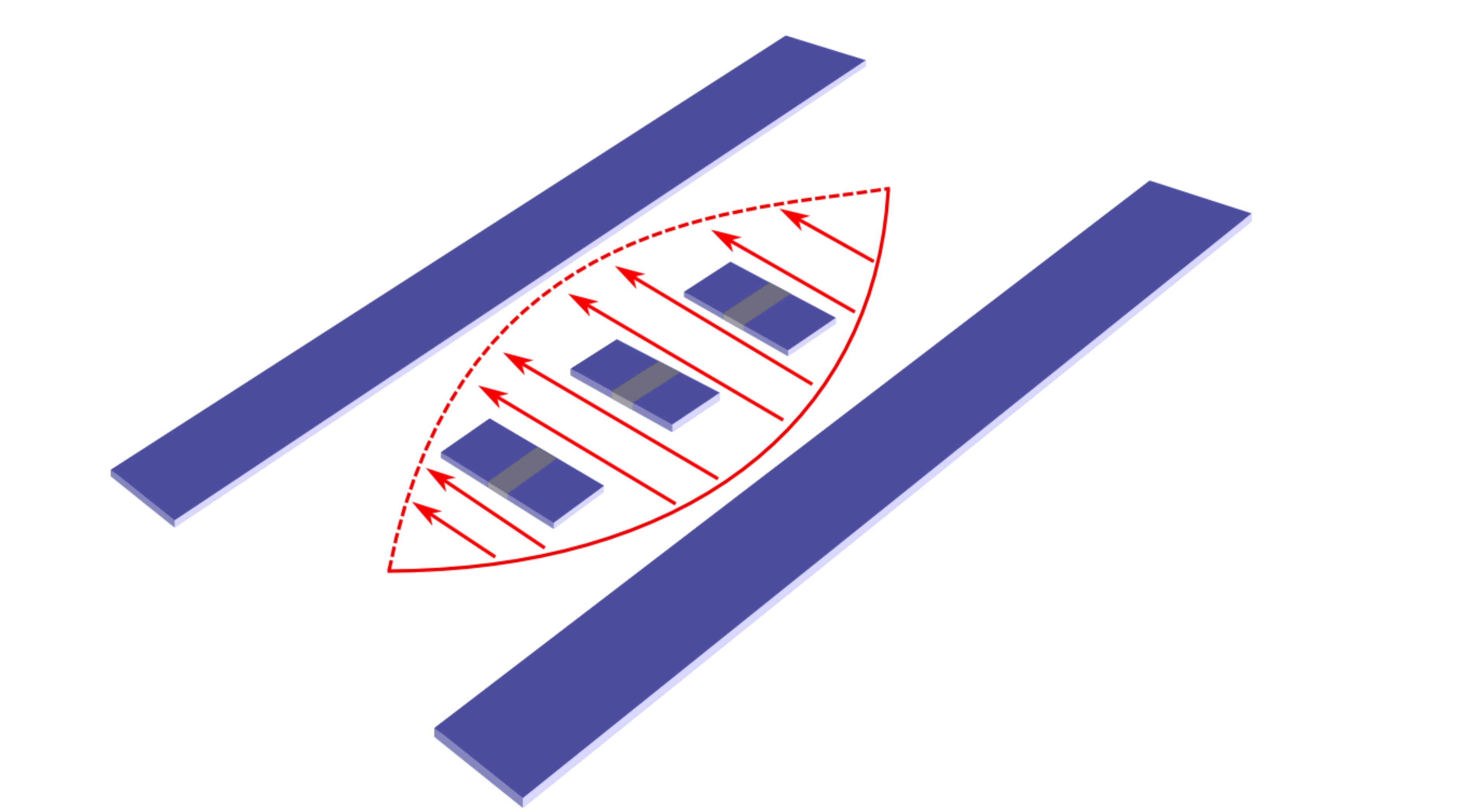}}
\caption{Schematic layout of a Cooper pair box array inside a microwave resonator of coplanar geometry with one-dimensional superconducting transmission line (stripline resonator), similar to proposed in \cite{Wallraff} for achieving of a strong coupling $g$ of two-level systems to the resonant photon in our frustrated Dicke Hamiltonian model.}
\label{1D_G}
\end{figure}

Combining together (\ref{gc}) and expression $g={2 el\sqrt{4\pi}}/(\sqrt{V})$, one may formulate the condition for the dipolar quantum phase transition as: 
\begin{align}
4 e l\sqrt{{\pi }/{ V}} \,= \sqrt{2N E_J} \, \label{omega_c},
\end{align}
where $l$ is a penetration depth of electric field into Cooper pair box superconducting island and $V$ is volume of the microwave cavity. Taking into account, that charging energy $E_C=\left(2 e\right)^2/2 C$ is of order: $E_C = {2 e^2}/{l}$, one may rewrite (\ref{omega_c}) in the following form:
\begin{align}
L\approx l\frac{E_Cl^2}{E_JN\Sigma}\,\label{L_l}\,,
\end{align}
where $\Sigma$ and $L$ are wave-guide (microwave cavity) cross-section area and length respectively. Assuming $L\approx Nl$ we finally find the following condition:
\begin{align}
N^2\approx \frac{E_Cl^2}{E_J\Sigma}\,\label{N_N}\,.
\end{align}
Hence, we come to a similar conclusion (see Figure \ref{1D_G}) as was already made in \cite{Wallraff}, that in order to achieve strong coupling limit $g\ge g_\mathrm{c}$ for a Cooper pair box array of a `thermodynamic size' $N\approx 10^2$ inside a microwave resonator, a coplanar geometry with one-dimensional superconducting transmission line (stripline resonator) should be used, thus providing inequality $S/l^2\ll 1$, and Cooper pair box should have charging energy much greater than Josephson coupling energy: $E_C\gg E_J$. 

\section{Conclusions}
     In summary, we have demonstrated that strong enough capacitive coupling of the Cooper pair box array of low-capacitance Josephson junctions  to a microwave resonant photon may lead to a first order quantum phase transition. As a result, a dipolar ordered state of Cooper pairs is formed, coupled to emerged coherent photonic condensate. The phase transition is of the first order due to infinitely coordinated antiferromagnetic (frustrating) interaction, that arises between Cooper pair dipoles of different Cooper pair boxes. This frustrating interaction is induced by the gauge-invariant coupling of the Josephson junctions to the resonant photon vector-potential in the microwave cavity. The strength of the coherent electromagnetic radiation field that emerges under the phase transition is proportional to the number $N$ of the Cooper pair boxes in the array and is reminiscent of the superradiant state of Dicke model without frustrating term found previously \cite{Brandes}. Nevertheless, the phase transition into the latter state is of second order \cite{Brandes} (see also Figure \ref{theta_app} in the Appendix \ref{app_Dicke}). 
 \newline
     The analytical description of the first order quantum phase transition in the Dicke model with infinitely coordinated antiferromagnetic frustrating interaction has become possible due to a new analytic tool: self-consistently `rotating' Holstein-Primakoff representation for the Cartesian components of the total spin, which is described in this paper. Our approach enables, as a by-product,  description of the second order quantum phase transition in the Dicke model without frustrating antiferromagnetic interaction, explored previously by other authors \cite{Brandes}. Nevertheless, `rotating' Holstein-Primakoff representation remains to be semiclassical  ($S\to \infty$). Therefore, the region of `spin liquid' with $S\sim 1$ is not attainable within this method.

\section {ACKNOWLEDGMENTS}
The authors acknowledge illuminating discussions with Carlo Beenakker, Konstantin Efetov and Bernard van Heck during the course of this work. This research was supported by the Netherlands Organization for Scientific Research (NWO/OCW), an ERC Synergy Grant, the Russian Ministry of Education and Science via the Increase Competitiveness Program of NUST MISiS grant No. K2-2017-085, and 'Goszadaniye' grant № 3.3360.2017/PH.

\begin{widetext}
\appendix

\section{Bogoliubov's transformation for the frustrated Hamiltonian} \label{app_Cooper pair box}
Below we show in detail a diagonalization procedure of the Hamiltonian (\ref{H_ab}).
Let's introduce 
\begin{align}
\hat{p}_x=\mathrm{i}\frac{1}{\sqrt{2\omega}}\left(\hat{a}^\dag-\hat{a}\right) \;\;\; \text{and} \;\;\; \hat{x}=\sqrt{\frac{\omega}{2}}\left(\hat{a}^\dag+\hat{a}\right) \,\label{xpx} 
\end{align}
together with
\begin{align}
\hat{p}_y=\mathrm{i}\frac{1}{\sqrt{2 E_J}}\left(\hat{b}^\dag-\hat{b}\right) \;\;\; \text{and} \;\;\; \hat{y}=\sqrt{\frac{E_J}{2}}\left(\hat{b}^\dag+\hat{b}\right) \,\label{ypy} 
\end{align}
and rewrite (\ref{H_ab}) in terms of (\ref{xpx}, \ref{ypy}):
\begin{align} \nonumber
&{}\hat{H}= - E_J \left(S+\frac{1}{2}\right) + \frac{1}{2}\left(\hat{x}^2 +\omega^2 \, \hat{p}_x^2\right) + \frac{1}{2}\left(\hat{y}^2 +E_J^2 \, \hat{p}_y^2\right) + \\{}&+\omega g \sqrt{S E_J} \, \hat{p}_x \hat{p}_y + \frac{g^2 S E_J}{2}\, \hat{p}_y^2 = - E_J \left(S+\frac{1}{2}\right) + \frac{1}{2}\hat{K}_{xy} + \frac{1}{2}\hat{K}_{p_xp_y}\, \label{Hxy},
\end{align}
where
\begin{align}
&{}\hat{K}_{xy} = \hat{x}^2+\hat{y}^2 \, \label{Kxy},\\
&{}\hat{K}_{p_xp_y} = \omega^2\hat{p}_x^2+E_J\left(E_J+g^2 S \right)\hat{p}_y^2 + 2 \,\omega g \sqrt{S E_J} \hat{p}_x \hat{p}_y\, \label{Kpxpy}.
\end{align}
We diagonalize (\ref{Hxy}) by performing a linear transformation of the quantum operators:
\begin{align}
\begin{pmatrix}
\hat{p}_x \\ \hat{p}_y
\end{pmatrix} = \begin{pmatrix}
\cos \gamma & \sin \gamma \\ -\sin \gamma & \cos \gamma
\end{pmatrix} \begin{pmatrix}
\hat{p}_1 \\ \hat{p}_2
\end{pmatrix} \;\;\; \text{and} \;\;\; \begin{pmatrix}
\hat{x} \\ \hat{y}
\end{pmatrix} = \begin{pmatrix}
\cos \gamma & \sin \gamma \\ -\sin \gamma & \cos \gamma
\end{pmatrix} \begin{pmatrix}
\hat{q}_1 \\ \hat{q}_2
\end{pmatrix}\, \label{Bg_transf}.
\end{align}
Then
\begin{align}
\hat{K}_{xy} = \hat{q}_1^2 + \hat{q}_2^2\, \label{Kq1q2},
\end{align}
and
\begin{align} \nonumber
\hat{K}_{p_xp_y}&{}= \left(\omega^2 \cos^2 \gamma + E_J \left(E_J +g^2 S\right) \sin^2 \gamma - 2 \omega g \sqrt{S E_J} \sin \gamma \cos \gamma\right)\hat{p}_1^2 + \\\nonumber&{}+
\left(\omega^2 \sin^2 \gamma + E_J \left(E_J +g^2 S\right) \cos^2 \gamma + 2 \omega g \sqrt{S E_J} \sin \gamma \cos \gamma\right)\hat{p}_2^2 + \\&{}+ \left( \left(\omega^2-E_J\left(E_J+g^2 S\right)\right)\sin 2\gamma + 2 \omega g \sqrt{S E_J} \cos 2\gamma\right) \hat{p}_1 \hat{p}_2\, \label{Kp1p2}.
\end{align}
The diagonalization condition that eliminates the cross-term $\sim \hat{p}_1 \hat{p}_2$, is:
\begin{align}
\tan 2\gamma = \frac{2 \omega g \sqrt{S E_J}}{E_J\left(E_J+g^2 S\right)-\omega^2} \, \label{tan}.
\end{align}
So, diagonalized operator $\hat{K}_{p_xp_y}$ becomes:
\begin{align}
\hat{K}_{p_xp_y} = \varepsilon_1^2 \, \hat{p}_1^2 + \varepsilon_2^2 \, \hat{p}_2^2\, \label{Kp1p2_diag}, 
\end{align}
where:
\begin{align} 
&{}2 \varepsilon_1^2=  E_J\left(E_J+g^2 S\right) + \omega^2  - \left(E_J\left(E_J+g^2 S\right)-\omega^2\right) \cos 2\gamma - 2 \omega g \sqrt{S E_J} \sin 2\gamma \, \label{eps12d}, \\
&{}2 \varepsilon_2^2= E_J\left(E_J+g^2 S\right) +\omega^2   + \left(E_J\left(E_J+g^2 S\right)-\omega^2\right) \cos 2\gamma + 2 \omega g \sqrt{S E_J} \sin 2\gamma \, \label{eps22d}.
\end{align}
Substitution of (\ref{tan}) to (\ref{eps12d}) and (\ref{eps22d}) gives the eigenvalues:
\begin{align}
2 \varepsilon_{1,2}^2 =  E_J\left(E_J+g^2 S\right) +\omega^2 \pm \sqrt{\left(E_J\left(E_J+g^2 S\right)-\omega^2\right)^2 + 4 \omega^2 g^2 S E_J} \, \label{eps_normal_ap2}. 
\end{align}
The transformation:
\begin{align}
\hat{p}_{1,2}=\mathrm{i} \frac{1}{\sqrt{2\varepsilon_{1,2}}} \left(\hat{c}^\dag_{1,2}-\hat{c}_{1,2}\right) \;\;\; \text{and} \;\;\; \hat{q}_{1,2} = \sqrt{\frac{\varepsilon_{1,2}}{2}} \left(\hat{c}^\dag_{1,2}+\hat{c}_{1,2}\right) \label{c1c2}
\end{align}
finally gives the diagonal Hamiltonian (\ref{H_diag}). 

The initial operators $\hat{a}$ and $\hat{b}$ are expressed via the new operators $\hat{c}^\dag_{1,2}$ as:
\begin{align}
&{}a = \sqrt{\frac{\omega}{\varepsilon_1}} \cos\gamma \; \hat{c}_1 + \sqrt{\frac{\omega}{\varepsilon_2}} \sin\gamma \; \hat{c}_2 \, \label{a}
\end{align}
and 
\begin{align}
&{}\hat{b} = -\sqrt{\frac{E_J}{\varepsilon_1}} \sin\gamma \; \hat{c}_1 + \sqrt{\frac{E_J}{\varepsilon_2}} \cos \gamma \; \hat{c}_2 \, \label{b},
\end{align}
where $\gamma$ is defined in (\ref{tan}).

\section{Second order quantum phase transition in the Dicke model within the RHP method} \label{app_Dicke}

We consider the standard Dicke Hamiltonian \cite{Dicke,Brandes} (modulo our notations)
\begin{align}
\hat{H}=\frac{1}{2}\left(\hat{p}^2+\omega^2 \hat{q}^2 \right) + g \hat{p} \, \hat{S}^y  - E_J \, \hat{S}^z   \, \label{H_Dicke_app}
\end{align}
at small coupling $g$.
We apply (\ref{S_z_HP}, \ref{S_y_HP}) to \ref{H_Dicke_app}: 
\begin{align} 
&{}\hat{H} = -E_J \left(S+\frac{1}{2}\right) +\omega\left( \hat{a}^\dag \hat{a}+\frac{1}{2}\right) + E_J\left( \hat{b}^\dag \hat{b}+\frac{1}{2}\right) - \frac{g\sqrt{S\omega}}{2}  \left(\hat{a}^\dag-\hat{a}\right) \left(\hat{b}^\dag  - \hat{b} \right) \, \label{H_ab_Dicke_app}. 
\end{align}
The Bogoliubov's transformation, similar to those in the Appendix \ref{app_Cooper pair box}, gives:
\begin{align}
\hat{H}=-E_J\left(S+\frac{1}{2}\right) + \varepsilon_1\left(\frac{1}{2}+ \hat{c}^\dag_1 \hat{c}_1\right)+ \varepsilon_2 \left(\frac{1}{2}+ \hat{c}^\dag_2 \hat{c}_2\right) \, \label{H_diag_Dicke_app},
\end{align}
with the excitations spectrum described by the new oscillator frequencies:
\begin{align}
2 \varepsilon_{1,2}^2 =  E_J^2 +\omega^2 \pm \sqrt{\left(E_J^2-\omega^2\right)^2 + 4 \omega^2 g^2 S E_J} \, \label{eps_normal_Dicke_app}. 
\end{align}
The ground state energy equals:
\begin{align}
E_0(S)=-E_J\left(S+\frac{1}{2}\right) + \frac{1}{2}\left(\varepsilon_1+\varepsilon_2\right) \, \label{E_0_Dicke_app}
\end{align}
One can check that as a function of $S$ the energy $E_0(S)$ has minimum at $S= N/2$, i.e. at the end of the interval of all possible total spin values $0\leq S\leq N/2$. This fact justifies the Holstein-Primakoff approach (\ref{S_z_HP}-\ref{S_x_HP}) valid in the large spin limit.

However, the lowest branch of excitations becomes imaginary when $ g> g_{\mathrm{c}}=\sqrt{E_J/S}$:
\begin{align}
\varepsilon_2 = \sqrt{\frac{E_J^2 +\omega^2}{2} - \frac{\sqrt{\left(E_J^2-\omega^2\right)^2 + 4 \omega^2 g^2 S E_J}}{2} }\, \label{e_min_Dicke_app}
\end{align}

Thus, the ground state described above is unstable in the interval $g >g_c$, compare \cite{Brandes}.

The method described in Section \ref{RHP} (\ref{J}-\ref{a_shift}) transforms the Hamiltonian (\ref{H_ab_Dicke_app}) into:
\begin{align}\nonumber
&{}\hat{H}=\omega\left(\hat{c}^\dag \hat{c} + \mathrm{i}\sqrt{\alpha} \left(\hat{c}^\dag-\hat{c}\right) +\alpha +\frac{1}{2} \right) - E_J \cos \theta \left(S-\hat{b}^\dag \hat{b}\right)+\\\nonumber &{}+E_J \sin \theta \, \mathrm{i} \sqrt{\frac{S}{2}} \left(\hat{b}^\dag -\hat{b}\right) - \frac{g\cos\theta \sqrt{S \omega}}{2} \left(\hat{c}^\dag-\hat{c}\right)\left(\hat{b}^\dag - \hat{b}\right) + g \cos \theta \sqrt{2 \omega \alpha}  \, \mathrm{i} \sqrt{\frac{S}{2}} \left(\hat{b}^\dag -\hat{b}\right) + \\ &{}+ g \sin\theta \, \mathrm{i} \sqrt{\frac{\omega}{2}} \left(\hat{c}^\dag-\hat{c}\right)\left(S -\left\langle \hat{b}^\dag \hat{b}\right\rangle  \right) + g \sin\theta \sqrt{2 \omega \alpha} \left(S-\hat{b}^\dag \hat{b}\right) \, \label{H1_shifted_Dicke_app}.
\end{align}
Here we have decoupled cubic in $\hat{c},\hat{b}$ operators terms in a mean-field approximation.
Conditions for vanishing of the linear terms $\propto$ $\left(\hat{c}^\dag-\hat{c}\right)$ and $\left(\hat{b}^\dag-\hat{b}\right)$  in the Hamiltonian (\ref{H1_shifted_Dicke_app}) are: 
\begin{align}
\sqrt{2 \omega \alpha}+g \sin \theta \, \left( S - \left\langle\hat{b}^\dag \hat{b}\right\rangle \right)=0 \, \label{c_lin_Dicke_app}\\
E_J \sin \theta + g \cos \theta \sqrt{2 \omega \alpha} =0 \, \label{d_lin_Dicke_app}
\end{align}

Solving the system of equations (\ref{c_lin_Dicke_app}) and (\ref{d_lin_Dicke_app}) we find: 
\begin{align}
&{}\cos \theta = \frac{E_J}{S g^2}\left(1-\frac{\left\langle\hat{b}^\dag \hat{b}\right\rangle}{S} \right)^{\,-1} \simeq \frac{E_J}{S g^2} \equiv \frac{g_\mathrm{c}^2}{g^2} \, \label{cos_Dicke_app}, \\ 
&{} \sqrt{\alpha} = - \frac{g S}{\sqrt{2 \omega}} \left(1-\frac{ \left\langle\hat{b}^\dag \hat{b}\right\rangle}{S} \right) \sin \theta \simeq \,\frac{g S}{\sqrt{2 \omega}} \sqrt{1-\frac{g_\mathrm{c}^4}{g^4}} \label{alpha05_Dicke_app},
\end{align}
where both the shift $\sqrt{\alpha}$ and rotation angle $\theta$ are non-zero when $g > g_\mathrm{c}$.  Thus,  using solutions (\ref{cos_Dicke_app}) and (\ref{alpha05_Dicke_app}) we obtain the initial Hamiltonian (\ref{H1_shifted_Dicke_app}) in the form similar to (\ref{H_ab_Dicke_app}), but renormalised with $\cos\theta$ coefficients :
\begin{align}\nonumber
&{}\hat{H}=\frac{E_J S}{2 \cos \theta}\left(1-\cos^2 \theta\right) -\frac{E_J}{\cos \theta} \left(S+\frac{1}{2} \right) +\omega\left(\hat{c}^\dag \hat{c} +\frac{1}{2} \right)+\frac{E_J}{\cos \theta} \left(\hat{b}^\dag \hat{b}+\frac{1}{2} \right) -\\&{}-\frac{g\cos\theta \sqrt{S\omega}}{2} \left(\hat{c}^\dag-\hat{c}\right)\left(\hat{b}^\dag - \hat{b}\right) \label{H1_Dicke_app}.
\end{align}
Next, we perform Bogoliubov's transformation that diagonalizes (\ref{H1_Dicke_app}), by performing a linear transform of Bose-operators $\hat{c},\,\hat{b}$ into Bose-operators $ \hat{e}_{1,2}$, and obtain: 
\begin{align}
\hat{H} = \frac{E_J S}{2 \cos \theta}\left(1-\cos^2 \theta\right) -  \frac{E_J}{\cos \theta} \left(S+\frac{1}{2}\right) +  \tilde{\varepsilon}_1\left(\frac{1}{2}+ \hat{e}^\dag_1 \hat{e}_1\right)+  \tilde{\varepsilon}_2 \left(\frac{1}{2}+ \hat{e}^\dag_2 \hat{e}_2\right) \label{H1_diag_Dicke_app}
\end{align}
with the eigenvalues: 
\begin{align}
2 \tilde{\varepsilon}_{1,2}^2 =  \frac{E_J^2}{\cos^2 \theta} +\omega^2 \pm \sqrt{\left(\frac{E_J^2}{\cos^2 \theta}-\omega^2\right)^2 + 4 \omega^2  E_J^2} \, \label{eps_sr_Dicke_app}, 
\end{align}
where both branches are now real for $g> g_{\mathrm{c}}=\sqrt{E_J/S}$ due to renormalisation of the coefficients with $\cos\theta$ factors. We have expressed in (\ref{eps_sr_Dicke_app}) the coupling constant $g$ via $\cos\theta$ using the self-consistency relation (\ref{cos_Dicke_app}).  
As is obvious from (\ref{cos_Dicke_app}) and (\ref{alpha05_Dicke_app}), both the shift $\sqrt{\alpha}$ and rotation angle $\theta$ progressively deviate from zero with increasing coupling strength $g$ in the interval $g > g_\mathrm{c}$, thus, providing a description of the new stable phase of the system.

The ground state energy of the system is now: 
\begin{align}
\widetilde{E}_0(S)=-  \frac{E_J}{2\cos \theta} \left(S+1\right) -\frac{E_J S}{2} \cos \theta + \frac{1}{2} \left(\tilde{\varepsilon}_1 + \tilde{\varepsilon}_2 \right)\, \label{E0_sr_Dicke},
\end{align}
which always has a minimum at the end of the spin interval, at $S=N/2$, thus justifying the Holstein-Primakoff approximation at finite angles $\theta$.

Thus, we found the second order phase transition that is manifested by a gradual rotation of the total spin expectation value in the $y-z$ plane by an angle $\theta$: 
\begin{align}
{\left\langle\hat{p}\right\rangle} = - Sg\sin \theta = \begin{cases} 0, \;\;\; g < g_\mathrm{c} \\ \mp Sg \sqrt{1-g_\mathrm{c}^4/g^4}, \;\;\; g \ge g_\mathrm{c} \end{cases} \, \label{p_av_pht_Dicke_app}
\end{align}
and
\begin{align}
{{\left\langle \hat{d} \right\rangle}} = 2eSl \sin \theta = \begin{cases} 0\,, \;\;\; g < g_\mathrm{c} \\ \pm 2elS \sqrt{1-g_\mathrm{c}^4/g^4}, \;\;\; g \ge g_\mathrm{c} \end{cases} \, \label{d_Dicke_app} 
\end{align}
where $g_\mathrm{c}=\sqrt{2 E_J/N}$ and $S=N/2$. The angle $\theta$, that describes the transition is plotted in the Figure \ref{theta_app}.

\begin{figure}[h!!]
\centerline{\includegraphics[width=0.5 \linewidth]{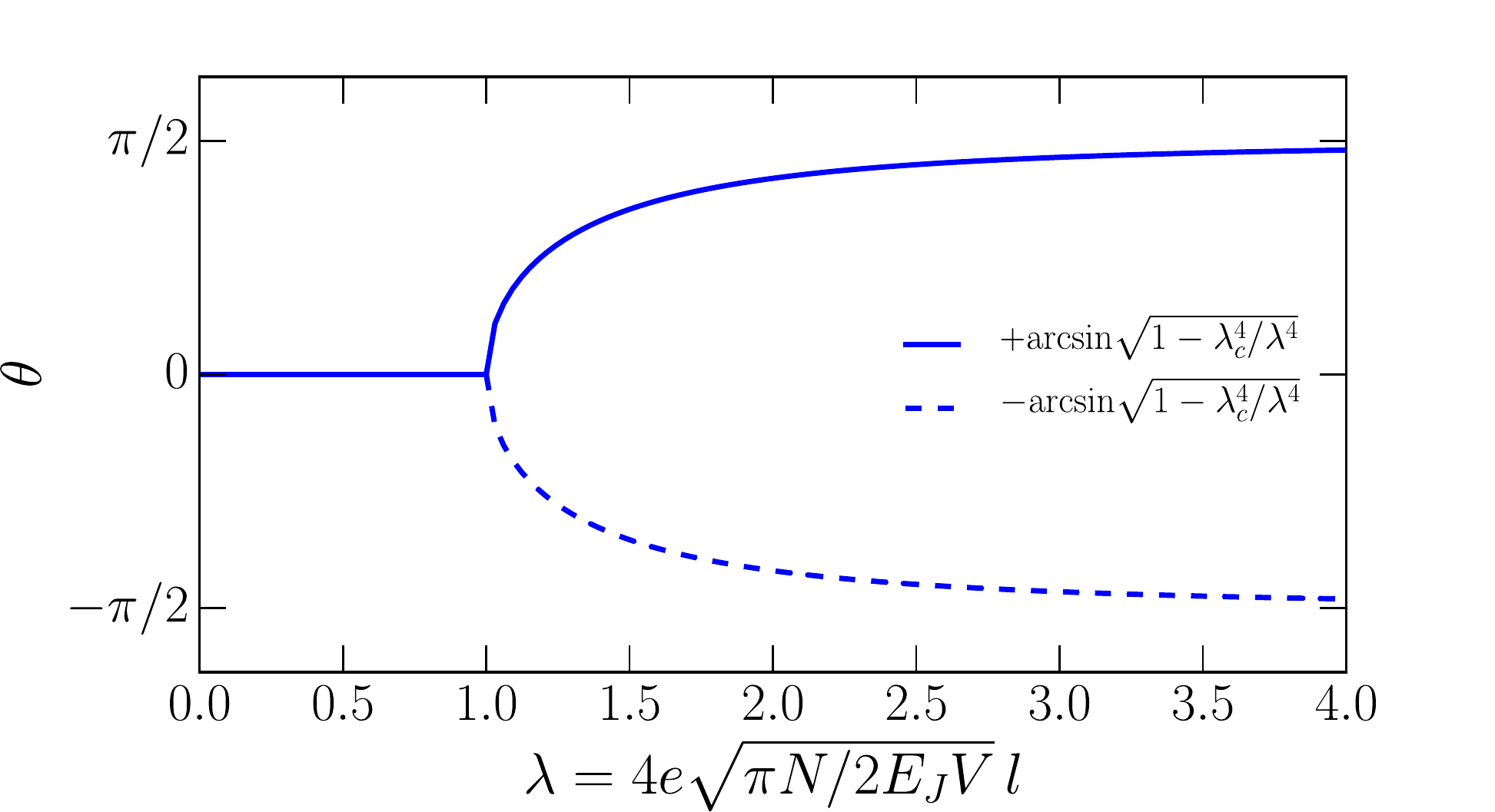}}
\caption{The angle $\theta$ as a function of dimensionless coupling constant $\lambda=g\sqrt{N/2E_J}$ in the Dicke model without frustration term. }
\label{theta_app}
\end{figure}
\end{widetext}

\end{document}